# Near-Visible Topological Edge States in a Silicon Nitride Platform

David Sharp,[1] Christopher Flower,[2] Mahmoud Jalali Mehrabad,[2] Arnab Manna,[1] Hannah Rarick,[1] Rui Chen,[3] Mohammad Hafezi,[2] and Arka Majumdar[1,3*]

[1]*Department of Physics, University of Washington, 3910 15th Ave NE, Seattle, WA 98195, USA*
[2]*Department of Electrical Engineering, University of Maryland, 8223 Paint Branch Dr, College Park, MD 20740, USA*
[3]*Department of Electrical and Computer Engineering, University of Washington, 185 E Stevens Way NE, Seattle, WA 91895, USA*
*\*arka@uw.edu*

**Abstract:** Demonstrations of topological photonics have so far largely been confined to infrared wavelengths where imaging technology and access to low-dimensional quantum materials are both limited. Here, we designed and fabricated silicon nitride ring-resonator arrays to demonstrate photonic topological edge states at ~780nm. We observed edge states corresponding to the integer quantum Hall Hamiltonian with topological protection against fabrication disorder. This demonstration extends the concept of topological edge states to the near-visible regime and paves the way for nonlinear and non-Hermitian topological photonics with the rich library of near-visible quantum emitters.

## 1. Introduction

Over the last two decades, topological photonics has emerged as an attractive platform to realize exotic physical models. Photonic analogues to the integer, spin, and valley quantum Hall effects as well as the Su-Schrieffer-Heeger model have all been demonstrated using arrays of ring resonators [1–3]. However, there are few demonstrations of such effects in photonic platforms operating at (near) visible wavelengths, primarily because they are often based on silicon photonics, which absorbs light of wavelengths below ~1000 nm. A few existing demonstrations in the visible regime have mostly been limited to topological photonic crystals and arrays of waveguides, which may not provide as versatile a platform as coupled ring resonators [4–6]. Implementing topological photonics in the (near) visible regime is attractive for two reasons. One, it is usually necessary to measure the spatial distribution of photons to fully characterize the topological nature of the system. In the infrared, this is hampered by poor imaging technology. In the (near) visible, one can employ silicon-based cameras with much higher resolution (larger number and smaller size of pixels). Two, (near) visible wavelength operation opens access to a large library of quantum emitters including two-dimensional semiconductors, colloidal semiconductor quantum dots, and thermal atomic vapors that are compatible with nanophotonics fabricated from CMOS-compatible materials [7–9]. Such emitters can extend the functionality of topological photonics by introducing single photon sources, optical gain, and large nonlinearities [10–15]. This would push topological photonics in new directions by expanding access to non-Hermitian and nonlinear topological regimes [16–18]. Furthermore, recent demonstrations of visible frequency combs [19-20] and infrared topological frequency combs [21-22] highlights the potential of topological photonics for harnessing rich microcomb physics in the (near) visible regime.

In this work, topological edge states at near-visible wavelengths were demonstrated in a system consisting of a two-dimensional lattice of coupled silicon nitride (SiN) ring-resonators. Like previous demonstrations, a photonic analogue to the integer quantum Hall (IQH) Hamiltonian was implemented by introducing a synthetic magnetic field for photons that mimics the effect of a uniform out-of-plane magnetic field on charged particles confined to a two-dimensional lattice [1,23]. Such a system in solid-state is expected to show edge states with topological protection against disorder [24]. Indeed, edge states were observed in the SiN ring-resonator lattice that were robust over a broad spectral range and against fabrication disorder. This system provides a powerful platform for further study of exotic materials systems as a photonic analogue in the near-visible regime. Importantly, this wavelength regime could enable low-power optical nonlinearity through integration of quantum emitters [25-26].

## 2. Design and Fabrication

The topological system in this work, depicted in Figure 1a, is a two-dimensional square lattice of ring-resonators coupled to each other via an interspersed array of detuned "link-ring" resonators. A carefully chosen shift in the position of certain link-rings introduces a position and direction dependent optical path length difference, simulating the direction-dependent phase produced by the magnetic field in the IQH model [1,27,28]. As a result, the photons in the lattice of ring-resonators experience a synthetic magnetic field and obey the IQH tight-binding Hamiltonian:

$$H = \sum_m \omega_0 a_m^\dagger a_m - \sum_{m,n} J_{m,n} \left( a_m^\dagger a_n e^{-i\varphi_{m,n}} + a_n^\dagger a_m e^{+i\varphi_{m,n}} \right) \quad (1)$$

where $a_m^\dagger$ is the photon creation operator at site $m = (m_x, m_y)$, $J_{m,n}$ is the hopping rate between sites $m, n$, which has non-zero value for only nearest-neighbor sites in the lattice, and $\varphi_{m,n}$ is the hopping phase between sites $m, n$ such that a photon traveling in a complete loop around a plaquette experiences a direction-dependent phase shift $\varphi = \pm\pi/2$.

The resulting spectrum consists of two edge bands separated by a bulk band. Each edge band has a set of topological edge states, confined to the edge of the lattice and propagating either clockwise (CW) or counterclockwise (CCW) around the lattice, depending on the band. These edge modes exhibit a linear dispersion relation in contrast with the modes of the bulk band, which are not spatially confined and do not have a well-defined momentum.

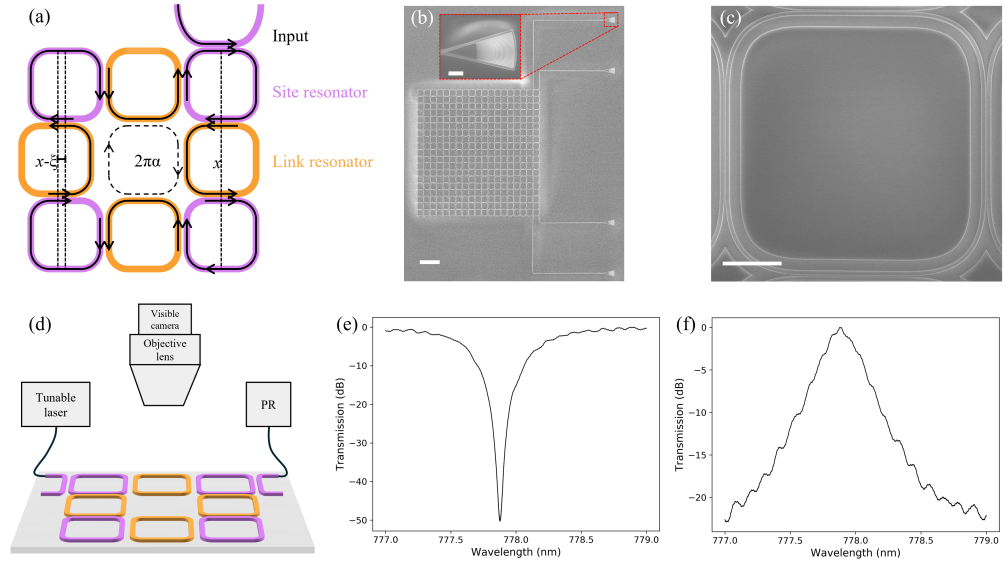

Fig. 1. (a) Schematic of a single plaquette where the site resonators and link resonators are denoted in purple and orange, respectively. The black arrows trace the path of light launched from the input waveguide and traveling clockwise around the lattice. A photon traveling in a complete clockwise loop around the plaquette accumulates a phase shift $2\pi\alpha$. Each column across the lattice is increasingly horizontally translated by $\zeta$ such that the phase shift is $\pi/2$. (b) Scanning electron micrograph (SEM) of an example 10x10 ring-resonator array. Light is coupled to and from the lattice using waveguides coupled to four separate apodized grating couplers. Inset: zoomed-in micrograph of an apodized grating coupler. The scale bars for the full-size image and the inset are 100 μm and 10 μm, respectively. (c) SEM of a site-ring situated in a larger array. The scale bar is 10 μm. (d) Schematic of the experimental set-up. A tunable laser launches light into the chip, and the transmission spectrum is measured by a photoreceiver (PR). The spatial distribution of photons is monitored by a 10x objective lens and visible camera. (e-f) Transmission drop (e) and through (f) spectra for a single representative site resonator used in the array.

The two-dimensional lattices of coupled ring-resonators were designed for 220 nm thick stoichiometric SiN on a $SiO_2$/Si substrate without any top cladding. The waveguide width was set to 600 nm such that the waveguides confine only a single transverse-electric mode at the target near-visible wavelength of 780 nm. This operating wavelength was chosen due to the lasers available and can be readily extended to the visible spectrum. The resonators were implemented in a racetrack geometry wherein coupling between adjacent resonators occurs over straight sections and the 90° bends occur at the corners. This geometry was chosen to ensure strong optical mode overlap between neighboring resonators. For the lattice in this work, the coupling length, coupling gap, and bending radius were chosen to be 12 μm, 175 nm, and 12 μm, respectively. Light was coupled to and from the chip using fully etched apodized grating couplers, which were designed to maximize the coupling efficiency while minimizing backscattering [29].

The pattern was transferred to the SiN using electron-beam lithography with ZEP520A positive-tone resist in conjunction with a low-pressure, low gas flow rate, fluorine-based plasma etch recipe designed to minimize sidewall roughness. All components of the devices, including grating couplers, were fully etched in the same step. Figure 1b depicts a scanning electron micrograph of an example 10×10 lattice. Light was coupled to and from the lattice via a pair of waveguides in an add-drop configuration that allowed for complete spectral characterization. On the same chip, a single site resonator associated with each two-dimensional lattice was also fabricated to characterize the component site-ring resonators, as depicted in Figure 1c.

## 3. Experimental Results

To characterize the ring-resonator lattice, an array of optical fibers was utilized in conjunction with a tunable-wavelength laser (New Focus TLB-6712-P) and photoreceiver (New Focus 2051-FS) (Figure 1d). This enabled measurement of the transmission spectra in both CW and CCW input configurations without altering the alignment of the chip relative to the excitation and collection pathways. In parallel, the intensity distribution of photons in the lattice was measured by monitoring scattering losses in the waveguides and ring-resonators on a CMOS camera (Allied Vision Prosilica GT 1930; 1936×1216 pixels) with a 10x (NA 0.28) objective lens. This experimental scheme was first used to characterize the extrinsic waveguide-resonator coupling rate and the intrinsic resonator decay rate for a single resonator $J \approx 52$ GHz and $\kappa_{in} \approx 9.4$ GHz, respectively (Figures 1e and 1f). From these measurements, the intrinsic quality-factor is estimated to be $Q \approx 40,000$.

Figure 2a depicts the transmission spectrum for the 8×8 ring-resonator lattice with CW site-ring input. The spectral ranges highlighted in purple and orange correspond to short and long edge states, respectively. These are evident in the intensity distribution of light scattered in the lattice (Figures 2b and 2c). Here, the short and long edge states are defined relative to the placements of the input waveguides and correspond to CW and CCW propagation around the lattice, respectively. The edge states propagate starting from the input waveguide at the top of the lattice and couple to the output waveguide at the bottom, which extinguishes the edge state. Two transmission peaks corresponding to two edge states are split by approximately $2J$ as expected for this system [24]. Each of these edge states are well-confined to the edges, and the long edge state tightly routes around each corner owing to topologically protected transport.

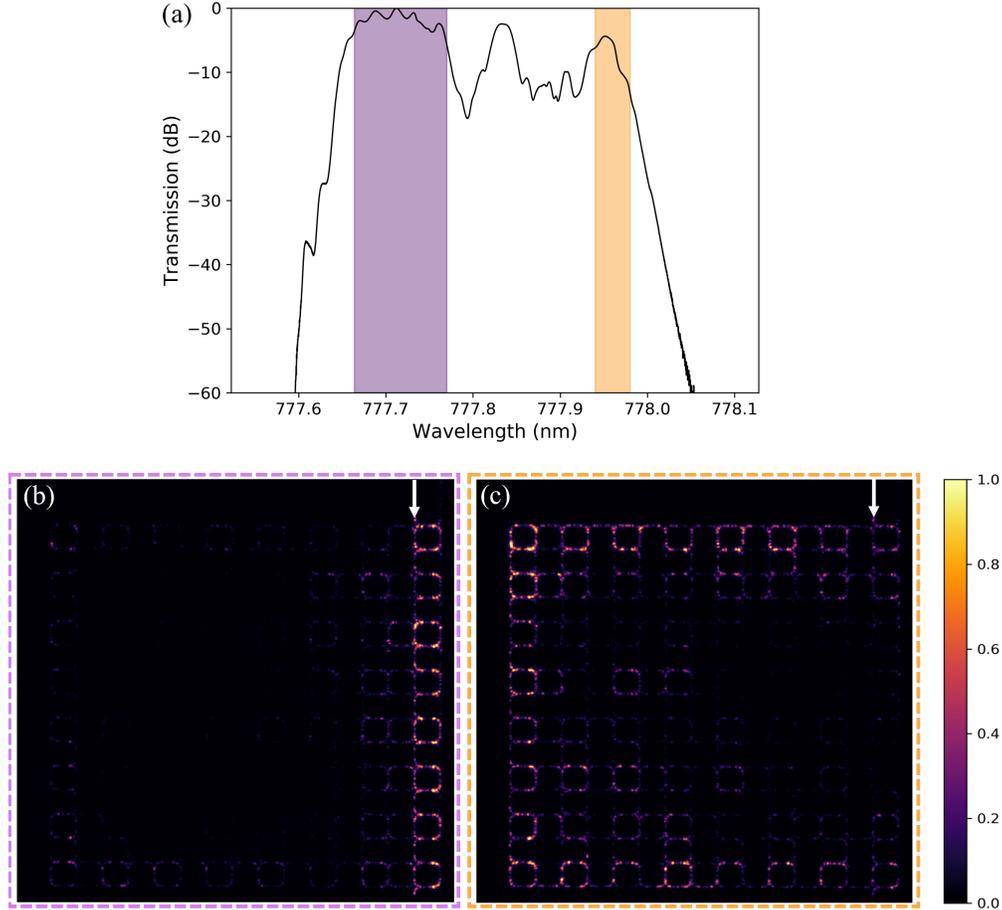

Fig. 2. (a) Measured transmission spectrum of the 8x8 ring-resonator array with CW input. The purple and orange regions highlight the approximate spectral regions of the short and long edge states, respectively. (b, c) Optical images of the spatial distribution of light scattered in the lattice at 777.720 nm (b) and 777.961 nm (c). Here, the short and long edges are defined with respect to the location of the input waveguide. (b-c) The white arrows indicate the locations of the inputs.

These measurements were repeated for the same spectral range but instead with CCW site-ring input. This corresponds to pumping the lattice with light of an opposite pseudo-spin. Since the synthetic magnetic field in this photonic system does not break time-reversal symmetry, this reverses the behavior as compared to CW site-ring input. This is evident in the transmission spectrum depicted in Figure 3a wherein the purple and orange highlighted regions still correspond to short and long edge states, respectively, but the spectral locations of each edge state are flipped relative to the central bulk state peak. The light again tightly routes around each corner and remains close to the edges (Figures 3b and 3c). For both input directions, each of the edge states maintains a similarity of its intensity distribution over a much broader bandwidth as compared to the bulk state (see Visualizations 1 and 2).

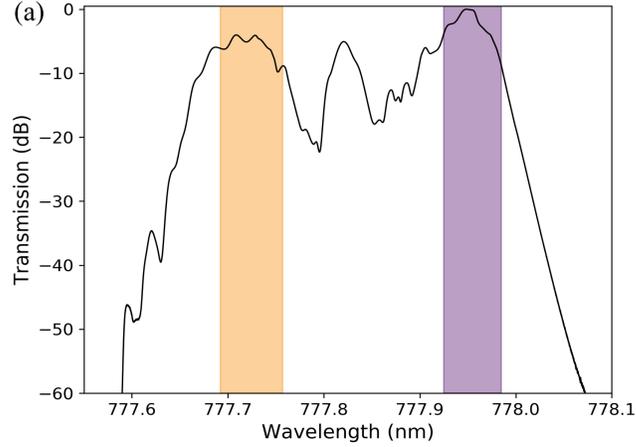

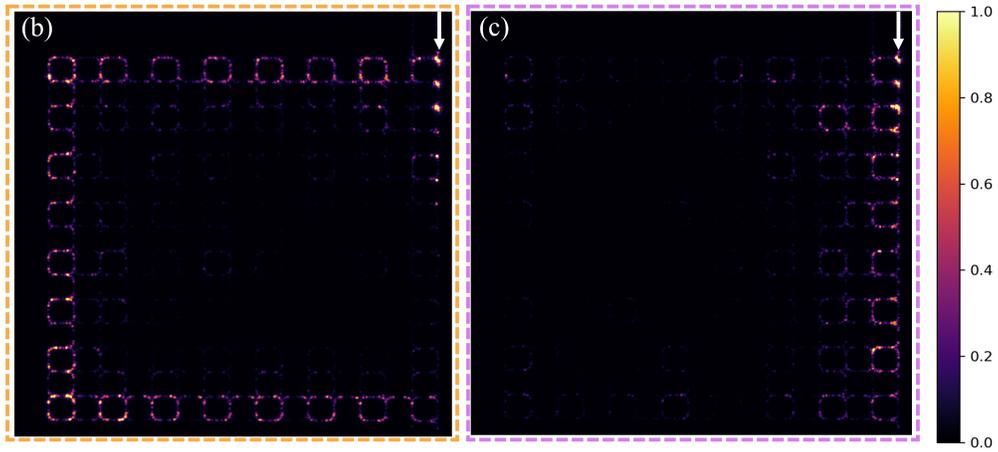

Fig. 3. (a) Measured transmission spectrum of the 8x8 ring-resonator array with CCW input. The orange and purple regions highlight the approximate spectral regions of the long and short edge states, respectively. (b, c) Optical images of the spatial distribution of light scattered in the lattice at 777.739 nm (b) and 777.982 nm (c). (b-c) The white arrows indicate the locations of the inputs.

Additionally, an otherwise identical 8×8 ring-resonator array with an intentionally removed site resonator was fabricated on the same chip (Figure 4a). This is an extreme case of disorder and serves to illustrate the effect of topological protection in the edge states. Light was coupled into the ring-resonator lattice with CW site-ring input at the short edge band and tightly routed around the missing resonator with minimal scattering into the bulk over a bandwidth of ~71 pm (35 GHz) (Figures 4b-d). This bandwidth is comparable with the unperturbed short edge band with CW site-ring input in Figure 2 and further confirms that the edge states in this system enjoy topological protection against disorder.

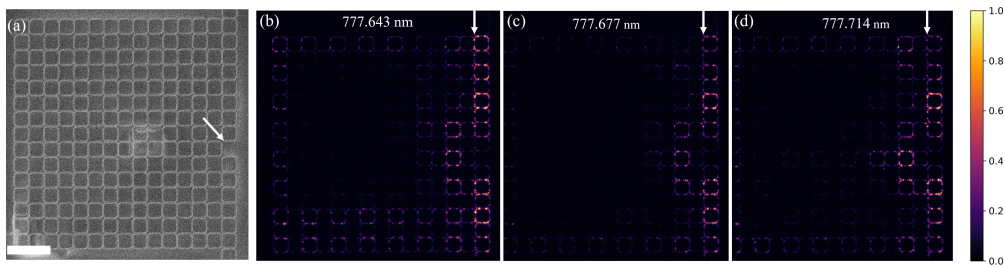

Fig. 4. (a) Scanning electron micrograph of the 8×8 ring-resonator array with an intentionally removed site resonator. The white arrow indicates the location of the missing resonator. The scale bar is 100 μm. (b-d) Optical images of the spatial distribution of light scattered in the lattice at 777.643 nm (b), 777.677 nm (c), and 777.714 nm (d). The light in the short edge state tightly rounds around the defect with minimal scattering into the bulk. (b-d) The white arrows indicate the locations of the inputs.

## 4. Conclusion

Topological edge states at near-visible wavelengths were demonstrated in a two-dimensional lattice of coupled SiN ring-resonators. This platform leveraged near-visible regime imaging technology to capture high-resolution images of the distribution of scattered light in the system, illustrating the imaging advantage of the near-visible regime. If the scattered light is coherent, one could use high-resolution (near) visible regime imaging to obtain the phase information of the photonic states in each resonator. This could enable phase measurements of the topological states, which have so far been elusive in any platform. Moreover, extension of topological photonics into near-visible wavelengths can enable visible wavelength topological frequency combs and paves the way for studies of many-body systems mimicked by topological photonics coupled to visible wavelength emitters. In particular, this latter effort could benefit greatly from recent advances in deterministic positioning of colloidal emitters [8,30,31].

**Funding.** This work is supported by the National Science Foundation under Grant No. DMR-2019444. Part of this work was conducted at the Washington Nanofabrication Facility/Molecular Analysis Facility, a National Nanotechnology Coordinated Infra-structure (NNCI) site at the University of Washington with partial support from the National Science Foundation via awards NNCI-542101 and NNCI-2025489.

**Disclosures.** The authors declare no conflicts of interest.

**Data Availability.** Data underlying the results presented in this paper are not publicly available at this time but may be obtained from the authors upon reasonable request.

**Supplemental document.** See Visualizations 1 and 2 for supporting content.